%% file: main.tex
\documentclass[twocolumn, conf]{IEEEtran}

\def\mode{4}  

\def\redactionmode{2} 

\def\alttextmode{1} 
\include{setups/setup}

\include{setups/acronyms}

\include{acronyms_local}

\definecolor{Gray}{gray}{0.9}
\definecolor{LightGray}{gray}{0.95}
\usepackage[table]{xcolor}

\usepackage{lipsum}
\usepackage{siunitx}
\usepackage{tabularray}
\usepackage{chemformula}
\usepackage[version=4]{mhchem}
\usepackage{amssymb}
\usepackage{pifont}

\UseTblrLibrary{diagbox}

\newcommand{\nor}{\overline{\lor}}

\begin{document}
\title{Cross-Layer Co-Design of Low-Latency Stateful Logic Adders with Printed Memristors}

\author{Fabian Seiler$^{\dagger,*}$, Kerem Kagan Güclü$^{+,*}$, Hongrong Hu$^+$, Gabriel Cadilha Marques$^+$,\\ Jasmin Aghassi-Hagmann$^+$, and Nima TaheriNejad$^\dagger$ \\
$^\dagger$Heidelberg University,  $^+$Karlsruhe Institute of Technology (KIT) \\
\{firstname.lastname\}@ziti.uni-heidelberg.de, \{firstname.lastname\}@kit.edu 
\\
$^*$These authors contributed equally
\thanks{
        This work was supported by the German Research Foundation (DFG) under Project 550835853 and Hector Stiftung.
        }
}

\maketitle

\begin{abstract}
Data movement between memory and processing units remains a major bottleneck in data-centric systems. Memristive \gls{sl} addresses this problem by performing Boolean operations directly on stored resistive states. However, arithmetic efficiency depends jointly on the device-supported logic primitive set, the dependency structure of the resulting operation sequence, and its mapping to datapath architectures. While prior designs typically optimize device, logic, and architecture layers separately, we jointly consider them to co-design a stateful adder.
We experimentally demonstrate a mixed MAGIC and FELIX logic primitive set on printed \gls{ecm} memristors and derive a full adder sequence optimized for inter-bit carry propagation. Mapping this sequence to bit-serial, bit-parallel, and bit-pipelined datapaths reduces the latter two to $2n+6$ cycles for an $n$-bit addition. 
For $32$-bit addition, this requires 60--88\% fewer cycles than prior \gls{sl} adders. Across four representative workloads, we achieve $3.7\times$--$16.4\times$ speedup over prior. These results show how cross-layer co-design can translate into improved workload-level performance.
\end{abstract}

\glsresetall

\section{Introduction}
\label{sec:1_intro}

Modern data-centric workloads are increasingly constrained by the energy and latency costs of moving data between memory and processing units~\cite{mutlu2025modernprimerprocessingmemory, MutluGoogleWorkloads}. \gls{imc} addresses this bottleneck by performing computation where data resides~\cite{Ielmini2018IMC_RRAM, lanza2025growing}. \gls{rram} devices are particularly attractive for \gls{imc}, as their non-volatile resistive states
provide both storage and computational functionality and enable parallel crossbar operations. While most memristive \gls{imc} systems target analog matrix--vector multiplication for neural-network acceleration~\cite{Sebastian2020MemoryDA}, digital memristive \gls{imc} instead employs \gls{sl} primitives to manipulate stored states directly. This enables deterministic Boolean and arithmetic operations while retaining data inside memory~\cite{borghetti2010nature, Kvatinsky_Review, Xu2021Review}.
Complex arithmetic in \gls{sl} is decomposed into sequences of dependent Boolean operations. Logic families such as IMPLY~\cite{borghetti2010nature}, MAGIC~\cite{Kvatinsky2014MAGIC}, and FELIX~\cite{Gupta2018Felix} expose different logic primitives, and therefore different arithmetic decompositions. 
Combining primitives across families is only useful if they are compatible with the same physical device, making the supported primitive set a key constraint on the resulting dependency graph~\cite{Hoffer2020NIMP, Seiler2026SoCC}. 
Addition is particularly sensitive to this interaction because its latency is often dominated by inter-bit carry propagation~\cite{Seiler2024NoCarry, RACER}. At the architectural level, \gls{sl} datapaths range from bit-serial execution~\cite{Talati2020mMPU} to partitioned and pipelined organizations with different opportunities for parallelism~\cite{AritPIM, RACER}. Consequently, adder efficiency depends not only on the number of operations, but also on how well its carry dependencies match the target datapath. 
Prior work typically either optimizes arithmetic sequences within a fixed logic family~\cite{Seiler2025CCMCC, SIMPLER, TaheriNejad2019SemiSerial, Rohani2020SemiParallel, talati2016nano, NanditMAGIC} or maps a fixed primitive set onto a target architecture~\cite{RACER, OSCAR, AritPIM}. 
This restricts both the available primitive set and the opportunity for datapath-aware sequence co-design. Here, we instead co-design the device-compatible primitive set, full adder sequence, and datapath mapping, and evaluate how these choices translate to workload-level performance.

The main contributions of this work are: 
i) experimental validation of a device-compatible mixed MAGIC and FELIX primitive set on printed \gls{ecm} memristors,
ii) a datapath-aware full adder algorithm with reduced inter-bit critical path,
iii) mappings to bit-serial, bit-parallel, and bit-pipelined datapaths, reducing latency down to $2n+6$ cycles for an $n$-bit addition, 
iv) a workload-level comparison showing $3.7\times$--$16.4\times$ speedup to comparable \gls{soa} adder.

\section{Physical Validation of Mixed Stateful Logic} \label{sec:2_primitives}
\subsection{Device Fabrication and Experimental Setup}

Printed logic circuits based on \gls{ecm} memristors were fabricated using laser ablation (Trumpf TruMicro 5000) for Au bottom electrode patterning and inkjet printing for the active layer deposition and the top electrode (Ag) deposition. Each memristor has an area of approximately $\SI{20}{\micro\meter}\times\SI{30}{\micro\meter}$. 

\begin{figure*}[t!]
    \centering
    \includegraphics[width=0.82\textwidth]{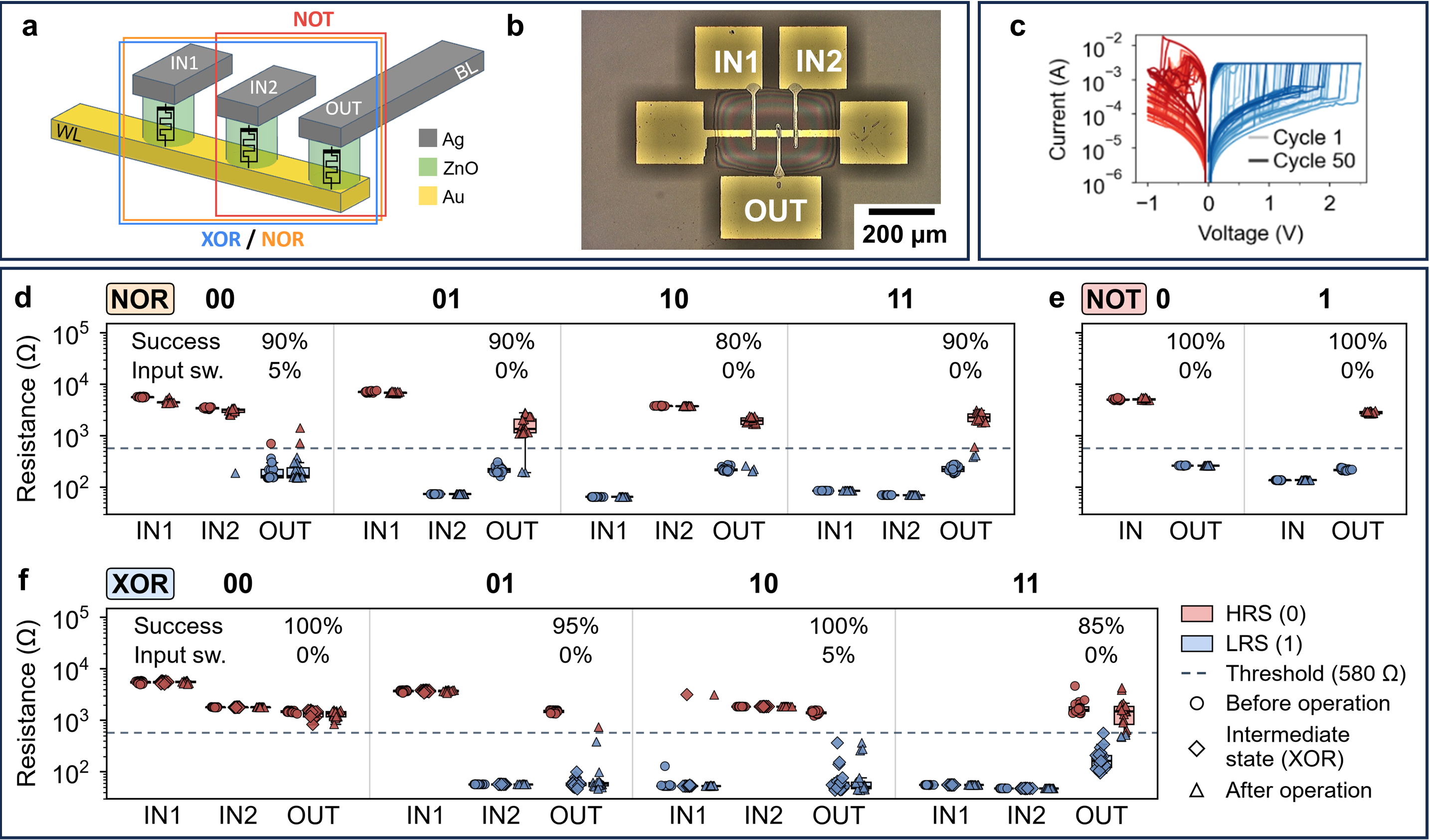}
    \vspace{-1mm}
    \caption{Experimental validation of the mixed stateful-logic primitive set on printed \gls{ecm} memristors. (a) Schematic of one crossbar row with three Ag/ZnO/Au memristors. The boxes indicate the memristors used for XOR/NOR and NOT. (b) Top-view optical image of the printed crossbar row. (c) $I$--$V$ curves of a representative device over 50 cycles. (d)--(f) Resistance of the input and output memristors for each input combination of (d) MAGIC NOR, (e) MAGIC NOT, and (f) FELIX XOR over 20 cycles.
    }
    \label{fig:experimental}
    \vspace{-3mm}
\end{figure*}

The $I$--$V$ curves of a representative device in~\Cref{fig:experimental}(c) show the typical bipolar resistive switching of the printed Ag/ZnO/Au \gls{ecm} memristors. The switching voltage and resistance state statistics have been analyzed in detail in our previous work~\cite{gulafshan2025realistic}, where the $V_\mathrm{SET}$ ($\mu_\mathrm{SET}=\SI{1.56}{\volt}$) is four times larger than the absolute value of $V_\mathrm{RESET}$ ($\mu_\mathrm{RESET}=\SI{-0.39}{\volt}$). 
This switching voltage asymmetry is crucial for the logic operations and is utilized as part of the cross-layer co-design~\cite{Seiler2026SoCC}.

MAGIC NOR, NOT, and FELIX XOR are crossbar compatible, which is demonstrated on one row of a crossbar as shown in~\Cref{fig:experimental}(a). The NOR and XOR circuits consist of three memristors (IN1, IN2, and OUT), whereas the NOT circuit consists of two memristors (IN1/IN2 and OUT). 
The pulse amplitudes for initialization and logic operations are listed in~\Cref{tab:parameters}, where each pulse has a width of 200 ms, while the SET and XOR,1 pulses have a 3 mA compliance current.

\vspace{-1.5mm}
\subsection{Validation of Stateful Logic Primitives}
 All operation voltages in~\Cref{tab:parameters} were determined experimentally by scanning the operation range, taking into account the boundary conditions proposed in~\cite{Kvatinsky2014MAGIC} and~\cite{Gupta2018Felix}. Since the upper boundary of each operation is set by the input cases in which the output memristor must retain its initialized state, the operation voltages were kept as high as possible without erroneous switching in these cases: $V_\mathrm{0,XOR1}$ without erroneous SET of the output in input case 00, $V_\mathrm{0,XOR2}$ without erroneous RESET of the output in input cases 01/10, and $V_\mathrm{0,NOR}$ and $V_\mathrm{0,NOT}$ without input switching in input cases 00 and 0, respectively. Each input combination was tested for 20 repetitions, and resistance states before and after operation are shown in~\Cref{fig:experimental}(d)--(f). The logic threshold of $R_\mathrm{th}=\SI{580}{\ohm}$ was calculated as the geometric mean of $\mu_\mathrm{HRS}$ ($\SI{1933}{\ohm}$) and $\mu_\mathrm{LRS}$ ($\SI{174}{\ohm}$), and is used to map \gls{lrs} to logic 1 and \gls{hrs} to logic 0.

\begin{table}[!t]
  \centering
  \caption{Pulse amplitudes for the utilized \gls{sl} operations.}
  \vspace{-1.5mm}
  \label{tab:parameters}
  \footnotesize
  \begin{tblr}{
    colspec  = {*{7}{Q[c,m]}},
    colsep   = 2.1pt,
    hline{1,Z} = {0.08em},
    hline{2}   = {0.05em},
  }
    \textbf{Operation}    & $V_\mathrm{SET}$ & $V_\mathrm{RESET}$ & $V_\mathrm{0,NOR}$ & $V_\mathrm{0,XOR1}$ & $V_\mathrm{XOR2}$ & $V_\mathrm{0,NOT}$ \\
    \textbf{Amplitude} (\si{\volt})     & \num{3.0} & \num{-1.5} & \num{1.15} & \num{2.6} & \num{-0.85} & \num{1.45} \\
  \end{tblr}
  \vspace{-5mm}
\end{table}

In MAGIC NOR and NOT, the output memristor is initialized to \gls{lrs}, and $V_\mathrm{0,NOR}$ or $V_\mathrm{0,NOT}$ is applied to the bit line (BL) of the each input memristor, while the BL of the output is grounded, so that the output is RESET to \gls{hrs} as long as one input is in \gls{lrs}. In FELIX XOR, the output is initialized to \gls{hrs} and the operation is performed in two cycles while grounding the BL of each input memristor: $V_\mathrm{0,XOR1}$ is applied to the BL of the output memristor to SET it if at least one input is in \gls{lrs} (OR), and $V_\mathrm{0,XOR2}$ subsequently RESETs the output only when both inputs are in \gls{lrs}.
 
As shown in~\Cref{fig:experimental}(d)--(f), the NOR, NOT, and XOR circuits achieved overall success rates of \SI{87.5}{\percent}, \SI{100}{\percent}, and \SI{95.0}{\percent}, respectively, where the failures mainly resulted from the output failing to RESET, particularly in input case 11 of XOR and input cases 01/10/11 of NOR. Input switching was observed in only \SI{1.25}{\percent} of the tests for NOR and XOR, and not at all for NOT. In NOR and NOT, this input stability, which is difficult to achieve when implementing MAGIC NOR with Valence Change Mechanism (VCM) memristors~\cite{Hoffer2020NIMP}, can be attributed to the asymmetric switching voltages of the printed \gls{ecm} memristors. As NOT and NOR are based on RESET of the OUT, the operation voltages are not large enough to switch the \gls{ecm} input memristors with a $|V_\mathrm{SET}$/$V_\mathrm{RESET}|$ of 4.0.

Overall, the low input switching of MAGIC NOR (\SI{1.25}{\percent}) and the error-free operation of MAGIC NOT constitute the main results of this experimental validation. Although the success rates of NOR and XOR are limited by device variability, the presented results are intended as a proof of concept that requires further device optimization. To the best of our knowledge, this is the first multi-cycle experimental demonstration of FELIX XOR and MAGIC NOR in the same memristor stack, which enables the cross-layer co-design of the carry-optimized full adder.

\begin{figure*}[t!]
    \centering
    \includegraphics[width=1\textwidth]{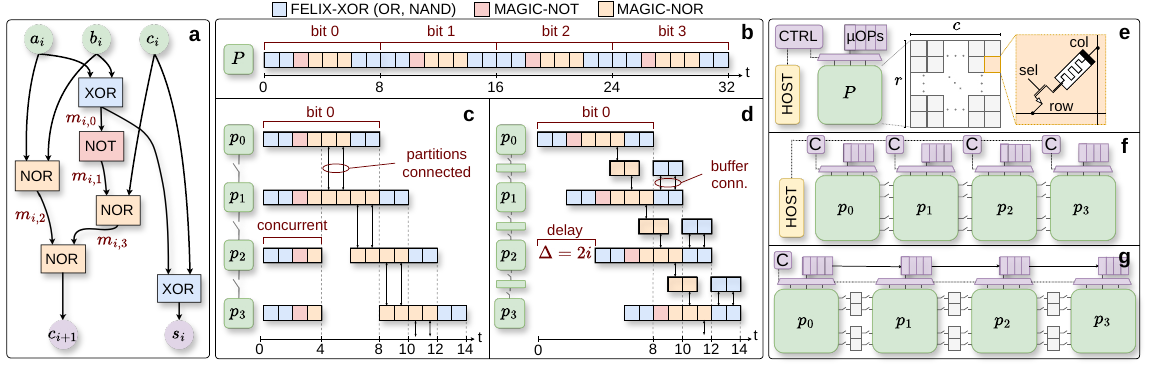}
    \vspace{-6mm}
    \caption{Proposed adder and datapath mappings. (a) Logic decomposition of a 1-bit full adder into stateful primitives. (b--d) Execution of four bit positions on serial, parallel, and pipelined datapaths. $P$ denotes a computational region, while $p_i$ denotes the partition for bit $i$. The x-axis indicates execution cycles. (e--g) Corresponding hardware organizations and control periphery.}
    \label{fig:datapath}
    \vspace{-3mm}
\end{figure*}

\section{Proposed Cross-Layer Mixed-Logic Adder} \label{sec:3_adder}

\subsection{Carry-Optimized Full Adder Design} 

The physical  demonstrated operations are abstracted here as single-cycle \glspl{uop}, hiding the underlying pulse conditions, device dynamics, and open reliability issues. The resulting set is $\{\texttt{NOR},\texttt{NOT},\texttt{OR},\texttt{NAND}\}$. Their composition provides the functional set $\{\oplus,\neg,\nor\}$, with XOR implemented in two cycles by consecutive \texttt{OR} and \texttt{NAND} \glspl{uop} using $x\oplus y \equiv (x\lor y)\land\neg(x\land y)$.
A full adder can then be expressed as
\begin{align}
    \mathrm{Sum}(a,b,c) &= a\oplus b\oplus c, \\
    \mathrm{Cout}(a,b,c) &= \left[(a\oplus b)\land c\right] \lor (a\land b) 
    \label{eq:fa}
\end{align}
The decomposition into sequentially arranged \glspl{uop} is shown in \Cref{fig:datapath}(a). The intermediate storage memristors are indicated by $m_{i,j}$ for bit $i$ and memristor index $j$.
Instead of simply optimizing cycle count, as in~\cite{NanditMAGIC, Seiler2025CCMCC, Gupta2018Felix, Hoffer2020NIMP}, we deliberately focus on minimizing the inter-bit carry propagation. The corresponding sequence of stateful primitives is listed in \Cref{tab:Algo}. Note that the updated state also depends on the initialization of the output memristor.
Its structure is chosen such that Steps~1--4 depend only on $a_i$ and $b_i$, while the inter-bit dependency is restricted to the two \texttt{NOR} operations in Steps~5--6. The final \texttt{XOR} in Steps~7--8 completes the sum after the incoming carry has been consumed. 
Consequently, the complete bit computation requires eight stateful \glspl{uop} (and two cycles required for initialization), whereas the carry critical path has a depth of only two. The required complemented states are therefore generated before the carry arrives. This distinction between local pre-computation and carry-dependent computation is the basis for the considered datapath mappings.

\begin{table}[ht!]
    \centering
    \caption{Stateful implementation of the proposed full adder.}
    \vspace{-1.5mm}
    \label{tab:Algo}
    \resizebox{\columnwidth}{!}{
    \begin{tabular}{c l l}
        \hline
        \textbf{Step} & \textbf{Micro-Operation} & \textbf{Equivalent Logic} \\
        \hline\hline
        \rowcolor{LightGray} Init & \multicolumn{2}{l}{$\mathrm{RESET}(m_{i,0}, s_i)$} \\
        \rowcolor{LightGray} Init & \multicolumn{2}{l}{$\mathrm{SET}(m_{i,1},m_{i,2},m_{i,3},c_{i+1})\quad$} \\
        
        1 & $m_{i,0} \leftarrow \mathrm{OR}(a_i,b_i)$
          & $m_{i,0} = a_i \lor b_i$ \\

        2 & $m_{i,0} \leftarrow \mathrm{NAND}(a_i,b_i)$
          & $m_{i,0} = a_i \oplus b_i$ \\

        3 & $m_{i,1} \leftarrow \mathrm{NOT}(m_{i,0})$
          & $m_{i,1} = \neg (a_i \oplus b_i)$ \\

        4 & $m_{i,2} \leftarrow \mathrm{NOR}(a_i,b_i)$
          & $m_{i,2} = a_i \nor b_i$ \\

        5 & $m_{i,3} \leftarrow \mathrm{NOR}(m_{i,1},c_i)$
          & $m_{i,3} = \neg (a_i \oplus b_i) \nor c_i$ \\

        \rowcolor{Gray}
        6 & $c_{i+1} \leftarrow \mathrm{NOR}(m_{i,2}, m_{i,3})$
          & $c_{i+1}=\mathrm{Cout}(a_i,b_i,c_i)$ 
          \\

        7 & $s_i \leftarrow \mathrm{OR}(m_{i,0},c_i)$
          & $s_i = (a_i\oplus b_i) \lor c_i$ \\

        \rowcolor{Gray}
        8 & $s_i \leftarrow \mathrm{NAND}(m_{i,0},c_i)$
          & $s_i = \mathrm{Sum}(a_i,b_i,c_i)$ 
          \\
        \hline
    \end{tabular}
    }
    \vspace{-4mm}
\end{table}

\subsection{Datapath-Aware Mapping}
\label{sec:datapath}

The proposed full adder sequence is mapped to bit-serial, bit-parallel, and bit-pipelined datapaths, as illustrated in \Cref{fig:datapath}(b--d). These organizations expose different degrees of spatial and temporal parallelism and therefore exploit the shortened carry dependency differently. All reported latencies are in cycles to allow technology-agnostic comparison, and exclude the two pre-initialization cycles.

\textbf{Bit-serial} datapaths reuse a single computational region $P$ across all bit positions, minimizing architectural overhead and commonly being used for \gls{sl} arithmetic~\cite{Seiler2025CCMCC, TaheriNejad2019SemiSerial, SIMPLER, Gupta2018Felix, Hoffer2020NIMP}. Since only one bit is processed at a time, this organization cannot exploit the shortened inter-bit dependency of the proposed algorithm. The complete eight-step sequence is therefore executed for bit $i$ before continuing with bit $i+1$, resulting in a latency of $L_{\mathrm{S}}=8n$ cycles for an $n$-bit addition.

\textbf{Bit-parallel} datapaths assign different bit positions $i$ to separate partitions $p_i$, allowing independent \glspl{uop} to execute concurrently~\cite{AritPIM, Oleitersdorf2022PartitionPIM}. For the proposed algorithm, execution is divided into two parts. First, steps~1--4 are broadcast and evaluated concurrently across all partitions, after which Steps~5--8 are staggered across consecutive bit positions. Since successive bits need to be displaced by only two cycles, the resulting latency is $L_{\mathrm{P}}=2n+6$.

\textbf{Bit-pipelined} organizations similarly assign one bit position $i$ to each partition $p_i$ and connect neighboring partitions through $r\times1$ memristive buffers~\cite{RACER, OSCAR}. Instead of independent partition schedules, a common instruction sequence is shifted by a delay between adjacent partitions, trading scheduling flexibility for regular pipelined execution. The carry-producing operation directly writes the inter-partition buffer, which is subsequently consumed by the following partition, avoiding dedicated buffer-transfer operations\footnote{RACER~\cite{RACER} uses a dedicated logic-0 column for $\texttt{NOR}(0,x)$ buffer transfers, whereas our buffer is connected to the general column-selection network and can directly participate in regular operations.}.
This yields a latency of $L_{\mathrm{PL}}=2n+6$ cycles.

\vspace{-1.5mm}
\subsection{Control and Hardware Trade-offs}
\label{sec:control}

The three datapaths use similar peripheral components, but differ in how computation is distributed and controlled~\cite{Seiler2026SoCC}, as shown in \Cref{fig:datapath}(e--g). Bit-serial execution uses a single computational region and instruction stream, resulting in the lowest hardware and control overhead, but cannot exploit the parallelism of the proposed mapping.
Bit-parallel execution adds $X=r(p-1)$ partition switches for $r$ rows and $p$ partitions and allows partitions to be scheduled independently. This provides the highest execution flexibility, but also requires the most complex control~\cite{Oleitersdorf2022PartitionPIM}. Bit-pipelined execution instead adds $2X$ switches and $X$ buffer memristors, while all partitions follow the same temporally shifted instruction stream~\cite{RACER}. It therefore trades additional communication hardware for more regular control and enables consecutive additions to overlap in the pipeline. These trade-offs are summarized in \Cref{tab:datapaths}.

\begin{table}[t]
    \centering
    \caption{Architectural trade-offs for datapath mappings.}
    \vspace{-1.5mm}
    \label{tab:datapaths}
    \resizebox{\columnwidth}{!}{
    \begin{tabular}{lccc}
        \hline
        & \textbf{Bit-serial}
        & \textbf{Bit-parallel}
        & \textbf{Bit-pipelined} \\
        \hline\hline

        \textbf{Structure}
        & Single array
        & $p$ partitions
        & $p$ partitions + buffers \\

        \multirow{1}{*}{\textbf{Extra HW}}
        & \multirow{1}{*}{--}
        & \multirow{1}{*}{$X$ switches$^+$}
        & $2X$ switches + $X$ memristors \\

        \textbf{Control (qual.)}
        & Low
        & High
        & Medium \\
        \hline
        
    \end{tabular}}
    \begin{tablenotes}
        \item \footnotesize $^+$$X$ indicates rows times partition interfaces $X=r(p-1)$.
    \end{tablenotes}
    \vspace{-2mm}
\end{table}

\section{Evaluation and Comparison}

\subsection{Adder-Level Comparison}
\label{sec:circuit}

\Cref{table:exact_vs_approx} compares the proposed mappings with \gls{soa} stateful adders in terms of compute cycles and required memristors. Pre-initialization is excluded, and row-wise \gls{simd} compatibility is reported since it determines whether independent additions can execute concurrently across crossbar rows.
The proposed parallel and pipelined mappings require only $2n+6$ cycles, or 70 cycles for $n=32$, while retaining row-wise \gls{simd} execution. This corresponds to approximately $60$--$88\%$ fewer cycles than the considered \gls{simd}-compatible baselines. The latency reduction comes at the cost of spatially replicating the bit-level computation, illustrating the trade-off established in \Cref{sec:datapath} and \Cref{sec:control}. Some transposed and semi-parallel designs achieve competitive cycle or device counts, but do not preserve the same row-wise \gls{simd} execution model and are therefore less suitable for the following throughput-oriented comparison.

\begin{table}[tb]
\setlength\extrarowheight{1pt}
	\centering
	\caption{Comparison of proposed $n$-bit adders with \gls{soa}.}
    \vspace{-1.5mm}
    \resizebox{1\columnwidth}{!}{
	\begin{tabular}{lccccccc}
		\hline
		\multirow{2}{*}{\textbf{Design}}  & \multirow{2}{*}{\textbf{\gls{simd}}} &
		\multicolumn{2}{c}{\textbf{Compute Cycles}$^\dagger$} & \multicolumn{2}{c}{\textbf{Num. Memristors}} \\
        &  & $n$-bit & $n=32$ & $n$-bit & $n=32$ \\
        \hline\hline
        IMPLY Serial~\cite{Seiler2025CCMCC} & Yes & $18n-1$ & 575 & $2n+4$ & 68 \\
        MAGIC Serial~\cite{SIMPLER} & Yes & $12n+1$ & 385 & $14n+1$ & 449 \\
        FELIX Serial~\cite{Gupta2018Felix} & Yes & $6n$ & 192 & $7n$ & 224 \\
        NIMP-OR Serial~\cite{Hoffer2020NIMP} & Yes & $7n+1$ & 225 & $6n+1$ & 193 \\ \hline
        
        IMPLY Parallel \cite{Karimi2018MemFA} & Yes & $5n+15$ & 175 & $4n+1$ & 129 \\
        MAGIC Pipelined~\cite{RACER}$^\ddagger$ & Yes &  $11n$ & 352 & $10n$ & 320 \\
        FELIX Pipelined~\cite{OSCAR}$^\ddagger$ & Yes & $9n$ & 288 & $9n$ & 288 \\ \hline

        IMPLY Semi-Serial \cite{TaheriNejad2019SemiSerial} & No & $10n+1$ & 321 & $2n+6$ & 70 \\
        IMPLY Semi-Parallel \cite{Rohani2020SemiParallel} & No & $17n-1$ & 543 & $2n+3$ & 67 \\

        MAGIC Trans. I~\cite{talati2016nano} & No & $10n+2$ & 322 & $16n-1$ & 511 \\
        MAGIC Trans. II~\cite{NanditMAGIC} & No & $7n+4$ & 228 & $16n$ & 512 \\ 
        \hline

        \textbf{Proposed Serial}
        & Yes & $8n$ & 256 & $8n+1$ & 257 \\
        \textbf{Proposed Parallel}
        & Yes & $2n+6$ & 70 & $8n+1$ & 257 \\
        \textbf{Proposed Pipelined}
        & Yes & $2n+6$ & 70 & $8n+1$ & 257 \\
        \hline
	\end{tabular}
    }
    \begin{tablenotes}
        \item \footnotesize $^\dagger$ Pre-initialization steps are disregarded in this comparison.
        \item \footnotesize $^\ddagger$ Sequence reconstructed, since not directly reported by~\cite{RACER, OSCAR}.
    \end{tablenotes}
	\label{table:exact_vs_approx}
    \vspace{-4mm}
\end{table}

\vspace{-1.5mm}
\subsection{Workload-Level Throughput}
\label{sec:application}

\begin{figure}[t]
    \centering
    \includegraphics[width=\linewidth]{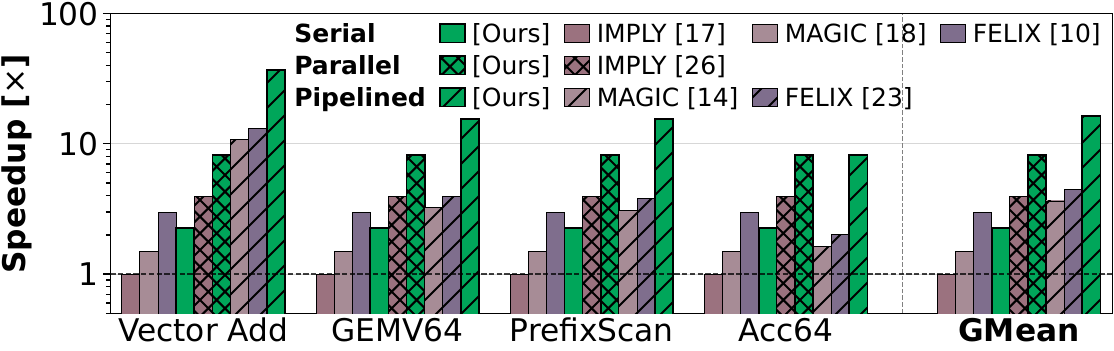}
    \vspace{-6mm}
    \caption{Workload speedup comparison to~\cite{Seiler2025CCMCC}. Colors distinguish logic families. Hatching denotes datapath organization. 
    }
    \label{fig:application_latency}
    \vspace{-4mm}
\end{figure}

To evaluate workload-level throughput, we use a custom cycle-accurate simulator in which workloads are represented as \glspl{dag}, with scalar additions as nodes and data dependencies as edges. A common greedy scheduler packs ready additions across the available \gls{simd} rows. A node becomes eligible once all of its predecessors have completed, while consecutive batches are issued according to the datapath-specific initiation interval. The resulting execution time therefore captures both single-add latency and sustained throughput under a common scheduling policy. Control and external data-transfer overheads are excluded for this comparison. Operands are assumed resident in the active tile, and intermediate results remain in memory.
All mappings use a $512\times512$ compute tile budget. Parallel and pipelined datapaths use 32 bit partitions, whereas the serial datapath uses a shared compute region. We evaluated four dependency structures that span throughput- to dependency-dominated execution: \textit{Vector Add} with 4096 independent additions, \textit{GEMV64} accumulates 64 independent 64-term reductions, \textit{PrefixScan} with a 1024-element stage-wise prefix scan, and \textit{Acc64} with 64 independent chains of 63 additions. 
\Cref{fig:application_latency} reports speedup $S=T_{\mathrm{ref}}/T$ relative to the serial IMPLY mapping~\cite{Seiler2025CCMCC} for representative works from all three datapaths~\cite{Seiler2025CCMCC, SIMPLER, Gupta2018Felix, Karimi2018MemFA, RACER, OSCAR}. 
The pipelined advantage is largest for \textit{Vector Add}, where independent additions fully expose inter-add overlap, and decreases as dependencies constrain concurrency. For \textit{Acc64}, each addition depends on the previous result, causing the proposed parallel and pipelined mappings to converge to the same latency. 
The proposed pipelined mapping achieves the highest speedup, improving over the proposed parallel mapping by approximately $2\times$ and over \gls{soa} mappings by $3.7\times$--$16.4\times$.
This shows that single-add latency alone does not determine workload performance. Sustained throughput also depends on the available dependency-level parallelism.

\section{Conclusion}
In this work, we presented a \gls{sl} adder developed through cross-layer co-design. We experimentally demonstrated MAGIC and FELIX primitives using the same \gls{ecm} device to derive a carry-optimized full adder sequence requiring only $2n+6$ cycles for an $n$-bit addition on parallel and pipelined datapaths. This reduces latency by $60\%$--$88\%$ compared to \gls{soa} \gls{sl} adders. For the evaluated workloads, our bit-pipelined approach exhibits a speedup of $3.7\times$--$16.4\times$ to the \gls{soa} baseline.
These results highlight the importance of jointly considering device, logic, and datapath architecture layers when designing \gls{sl} arithmetic.

\newpage
\IEEEtriggeratref{14}
\bibliographystyle{IEEEtran}
\bibliography{references}

\end{document}

%% file: setups/setup.tex
\usepackage{adjustbox}              
\usepackage{amsmath}
\usepackage{cite}					
\usepackage{cleveref}
\usepackage{comment}  				

\usepackage{enumitem}				
\usepackage{graphicx}
	\graphicspath{{./figures/}}		
\usepackage{multirow}				
\usepackage{tabularx} 				
\usepackage{threeparttable}
\usepackage{subfig}
\usepackage{soul} 					
\usepackage{xargs}    				
\usepackage[svgnames]{xcolor}			

\newcommand{\red}{}
\newcommand{\green}{}
\newcommand{\blue}{}

\newcounter{documentmode}
\includecomment{conf}  	
\newcommand{\setdocmode}{%
  \ifcase\number\value{documentmode}
	\renewcommand{\red}[1]{\color{red}##1\color{black}{}}
	\renewcommand{\green}[1]{\color{Green}##1\color{black}{}}
	\renewcommand{\blue}[1]{\color{blue}##1\color{black}{}}
  \or
    \usepackage{draftwatermark}
	\SetWatermarkScale{1.35}
	\SetWatermarkAngle{56}
	\SetWatermarkLightness{0.9}
	\SetWatermarkHorCenter{0.9\textwidth}
	\SetWatermarkVerCenter{0.7\textheight}
  \or
	\usepackage{draftwatermark}
	\SetWatermarkText{CONFIDENTIAL}
	\SetWatermarkScale{0.67}
	\SetWatermarkAngle{56}
	\SetWatermarkLightness{0.9}
	\SetWatermarkHorCenter{0.6\textwidth}
	\SetWatermarkVerCenter{0.6\textheight}
  \or
	\usepackage{draftwatermark}
	\SetWatermarkText{Pre-Publication}
	\SetWatermarkScale{2.67}
	\SetWatermarkAngle{56}
	\SetWatermarkLightness{0.9}
	\SetWatermarkHorCenter{0.6\textwidth}
	\SetWatermarkVerCenter{0.6\textheight}
  \else
    \excludecomment{conf}
  \fi
}

\if\mode0
\usepackage[prependcaption, textsize=footnotesize]{todonotes}
\else
\usepackage[disable, prependcaption, textsize=footnotesize]{todonotes}

\fi

\setdocmode

\newcounter{redmode}
\includecomment{2Detailed}			
\includecomment{Uncritical}			
\includecomment{2SaveSpace}			
\includecomment{IfMustDelete}			
\newcommand{\redact}{%
  \ifcase\number\value{redmode}
    \includecomment{2Detailed}
    \includecomment{Uncritical}
    \includecomment{2SaveSpace}
    \includecomment{IfMustDelete}
  \or
    \excludecomment{2Detailed}
    \includecomment{Uncritical}
    \includecomment{2SaveSpace}
    \includecomment{IfMustDelete}
  \or
    \excludecomment{2Detailed}
    \excludecomment{Uncritical}
    \includecomment{2SaveSpace}
    \includecomment{IfMustDelete}
  \or
    \excludecomment{2Detailed}
    \excludecomment{Uncritical}
    \excludecomment{2SaveSpace}
    \includecomment{IfMustDelete}
  \or
    \excludecomment{2Detailed}
    \excludecomment{Uncritical}
    \excludecomment{2SaveSpace}
    \excludecomment{IfMustDelete}
  \fi
}
\redact

\newcounter{alttxtmode}
\includecomment{TextA}			
\includecomment{TextB}			
\newcommand{\alttext}{%
  \ifcase\number\value{alttxtmode}
    \includecomment{TextA}
    \includecomment{TextB}
  \or
    \excludecomment{TextA}
    \includecomment{TextB}
  \or
    \includecomment{TextA}
    \excludecomment{TextB}
  \or
    \excludecomment{TextA}
    \excludecomment{TextB}
  \fi
}
\alttext

%% file: setups/acronyms.tex
\usepackage[acronym]{glossaries} 

\newacronym{am}{AM}{Amplitude Modulation}
\newacronym{asd}{ASD}{Autism Spectrum Disorder}
\newacronym{at}{AT}{Adaptive Tracking}
\newacronym{bw}{BW}{Baseline Wanderer}
\newacronym{bpm}{BPM}{Breath per Minute}
\newacronym{bvp}{BVP}{Blood Volume Pulse}
\newacronym{cemd}{CEMD}{Complex Empirical Mode Decomposition}
\newacronym{co}{CO}{Count Origin}
\newacronym{cosf}{TDCO}{Time Domain - Count Origin}
\newacronym{costf}{COSTF}{Count Origin - Smart and Time Fusion}
\newacronym{cwt}{CWT}{Continuous Wavelet Transform}
\newacronym{der}{DER}{Detection Error Rate}
\newacronym{df}{DF}{Dominant Frequency}
\newacronym{dfsf}{SFFDP}{Smart Fusion of Frequency Domain Peak}
\newacronym{dwt}{DWT}{Discrete Wavelet Transform}
\newacronym{ecg}{ECG}{Electrocardiography}
\newacronym{skin_eda}{EDA}{Electrodermal activity} 
\newacronym{eeg}{EEG}{Electroencephalogram}
\newacronym{emd}{EMD}{Empirical Mode Decomposition}
\newacronym{emg}{EMG}{Electromyography}
\newacronym{eog}{EOG}{Electrooculogram}
\newacronym{ews}{EWS}{Early Warning Score}
\newacronym{fft}{FFT}{Fast Fourier Transformation}
\newacronym{fm}{FM} {Frequency Modulation}
\newacronym{gsr}{GSR}{Galvanic Skin Response}
\newacronym{hf}{HF}{High Frequency}
\newacronym{hr}{HR}{Heart Rate}
\newacronym{hrv}{HRV}{Heart rate variability}
\newacronym{ibi}{IBI}{Inter beat interval}
\newacronym{ica}{ICA}{Independent Component Analysis}
\newacronym{iir}{IIR}{Infinite Impulse Response}
\newacronym{imf}{IMF}{Intrinsic Mode Function}
\newacronym{knn}{KNN}{k-nearest neighbour}
\newacronym{lda}{LDA}{Linear Discriminant Analysis}
\newacronym{led}{LED}{light emitting diode}
\newacronym{lms}{LMS}{Least Mean Square}
\newacronym{lf}{LF}{low frequency}
\newacronym{ma}{MA}{Movement Artifact}
\newacronym{mad}{MAD}{Mean Absolute Deviation} 
\newacronym{mar}{MAR}{Motion Artifact Removal} 
\newacronym{mimic}{MIMIC}{Multi-parameter Intelligent Monitoring for Intensive Care}
\newacronym{nlms}{NLMS}{Normalized Least Mean Square}
\newacronym{n2n}{N2N}{normal-to-normal}
\newacronym{pd}{PD}{Peak Detection}
\newacronym{pdsf}{TDPD}{Time Domain Peak Detection}
\newacronym{ppe}{PPE}{Peak to Peak Error}
\newacronym{ppg}{PPG}{Photoplethysmogram}
\newacronym{ptt}{PTT}{Pulse transit time}
\newacronym{rls}{RLS}{ Recursive Least Square}
\newacronym{rms}{RMS}{ Root Mean Square}
\newacronym{rmse}{RMSE}{Root Mean Square Error}
\newacronym{rr}{RR}{Respiratory Rate}
\newacronym{rrsv}{RRSV}{Repeated Random Sub-Sampling Validation}
\newacronym{sam}{SAM}{Self-Assessment Manikin}
\newacronym{saews}{SA-EWS}{Self-Aware Early Warning Score}
\newacronym{sart}{SART}{sustained attention to response test}
\newacronym{sfu}{SFU}{Smart Fusion}
\newacronym{scl}{SCL}{Skin Conductance Level}
\newacronym{scr}{SCR}{Skin Conductance Response}
\newacronym{sigvm}{SigVM}{Signal Vector Magnitude}
\newacronym{skt}{SKT}{Skin Temperature}
\newacronym{std}{STD}{Standard Deviation }
\newacronym{svd}{SVD}{Singular Value Decomposition }
\newacronym{tfu}{TFU}{Temporal Fusion } 
\newacronym{vfcdm}{VFCDM} {Variable Frequency Complex Demodulation}
\newacronym{whs}{WHS}{Wearable Health-care Systems}

\newacronym{ai}{AI}{Artificial Intelligence}
\newacronym{ann}{ANN}{Artificial Neural Network}
\newacronym{bpn}{BPN}{Back-Propagation Neural Network}
\newacronym{bpnn}{BPNN}{back-propagation neural network}
\newacronym{cart}{CART}{Classification And Regression Tree}
\newacronym{cnn}{CNN}{Convolutional Neural Network}
\newacronym{flop}{FLOP}{Floating Point Operation}
\newacronym{icnn}{ICNN}{Iterative Convolutional Neural Network}
\newacronym{ldf}{LDF}{Linear Discriminant Function}
\newacronym{mcs}{MCS}{Multiple Classifier System}
\newacronym{ml}{ML}{Machine Learning}
\newacronym{nn}{NN}{Neural Network}
\newacronym{nb}{NB}{Naive Bayesian}
\newacronym{rsvm}{RSVM}{Reputation-driven Support Vector Machine }
\newacronym{svm}{SVM}{Support Vector Machine}
\newacronym{ucnn}{$\mu$CNN}{Micro CNN}

\newacronym{brs}{BRS}{Bipolar Resistive Switch-based logic}
\newacronym{cnf}{CNF}{Conjunctive Normal Form}
\newacronym{crs}{CRS}{Complementary Resistive Switch-based logic}
\newacronym{dnf}{DNF}{Disjunctive Normal Form}
\newacronym{fpm}{FPM}{Forward Polarized Memristor}
\newacronym{hfo}{$HfO_x$}{Hafnium Oxide}
\newacronym{hrs}{HRS}{High Resistance State}
\newacronym{imc}{IMC}{In-Memory Computation}
\newacronym{imply}{IMPLY}{Material Implication}
\newacronym{imp}{IMP}{In-Memory Processing}
\newacronym{lim}{LIM}{Logic in Memory}
\newacronym{lrs}{LRS}{Low Resistance State}
\newacronym{ltg}{LTG}{Logic Threshold Gate}
\newacronym{magic}{MAGIC}{Memristor-Aided Logic}
\newacronym{mecoins}{Me-Coin}{Memristor-based Computation In-memory}
\newacronym{pcm}{PCM}{Phase Change Memory}
\newacronym{pim}{PIM}{Processing in Memory}
\newacronym{reram}{ReRAM}{Resistive Random Access Memory}
\newacronym{rpm}{RPM}{Reversely Polarized Memristor}
\newacronym{sdc}{SDC}{Self Directed Channel}
\newacronym{stt}{STT}{Spin Transfer Torque}
\newacronym{tao}{$TaO_x$}{Tantalum Oxide}
\newacronym{tio}{$TiO_2$}{Titanium dioxide}
\newacronym{vteam}{VTEAM}{Voltage-controlled ThrEshold Adaptive Memristor}

\newacronym{aco}{ACO}{Autonomous Cooperating Object}
\newacronym{afdd}{AFDD}{Automated Fault Detection and Diagnostic}
\newacronym{ca}{CA}{Continuous Average}
\newacronym{cah}{CAH}{Context-Aware Health Monitoring}
\newacronym{cam}{CCAM}{Confidence-based Context-Aware condition Monitoring}
\newacronym{csa}{CSA}{Computational Self-Awareness}
\newacronym[plural=DABs,longplural={Discrete Average Blocks}]{dab}{DAB}{Discrete Average Block}
\newacronym[plural=KPNs,longplural={Kahn Process Networks}]{kpn}{KPN}{Kahn Process Networks} 
\newacronym{mape-k}{MAPE-K}{Monitor-Analyze-Plan-Execute over a shared Knowledge}
\newacronym[plural=MoCs,longplural={Models of Computation}]{moc}{MoC}{Model of Computation}
\newacronym{oda}{ODA}{Observe-Decide-Act}
\newacronym{pca}{PCA}{Principal Component Analysis}
\newacronym{rosa}{RoSA}{Research on Self-Awareness}
\newacronym{sa}{SA}{Self-Aware}
\newacronym{saness}{SA}{Self-Awareness}
\newacronym{sahm}{SAHM}{Self-Aware Health Monitoring}
\newacronym{samba}{SAMBA}{Self-Aware health Monitoring and Bio-inspired coordination for distributed Automation systems}
\newacronym{selphys}{SelPhyS}{Self-aware cyber-Physical System}
\newacronym{sh}{SH}{State Handler}
\newacronym{som}{SOM}{Self-Organizing Map}

\newacronym{c-s}{CAS}{Compare-and-Swap}
\newacronym{cps}{CPS}{Cyber-Physical System}
\newacronym{cpps}{CPPS}{Cyber-Physical Production System}
\newacronym{dsr}{DSR}{Down-Sampling Rate}
\newacronym{dum}{DuM}{Device under Monitoring}
\newacronym{es}{ES}{Embedded System}
\newacronym{LD}{LD}{low-discrepancy}
\newacronym{mes}{MES}{Manufacturing Execution System}
\newacronym{sc}{SC}{Stochastic Computing}
\newacronym{sos}{SoS}{System of Systems}
\newacronym{suo}{SuO}{System under Observation}

\newacronym{abi}{ABI}{Application Binary Interface}
\newacronym{adc}{ADC}{Analog-to-Digital Converter}
\newacronym{aes}{AES}{Advanced Encryption Standard}
\newacronym{alu}{ALU}{Arithmetic Logic Unit}
\newacronym{api}{API}{Application Programming Interface}
\newacronym{asic}{ASIC}{Application Specific Integrated Circuit}
\newacronym{asoc}{ASOC}{Autonomic System-on-Chip platform}
\newacronym{axi}{AXI}{Advanced eXtensible Interface Bus}
\newacronym{bram}{BRAM}{Block Random Access Memory}
\newacronym{cdt}{CDT}{C/C++ Development Tooling}
\newacronym{clb}{CLB}{Configuarable Logic Block}
\newacronym{cmos}{CMOS}{Complementary Metal-Oxide Semiconductor}
\newacronym{cp}{CP}{Clock Pulse}
\newacronym{cpi}{CPI}{Cycles Per Instruction}
\newacronym{cpu}{CPU}{Central Processing Unit} 
\newacronym{cpsoc}{CPSoC}{Cyber-Physical System-on-Chip}
\newacronym{cu}{CU}{Compute Unit}
\newacronym{cuda}{CUDA}{Compute Unified Device Architecture}
\newacronym{dac}{DAC}{Digital to Analog Converter}
\newacronym{ddr3}{DDR3}{Double Data Rate}
\newacronym{dff}{DFF}{Data Flip-Flop}
\newacronym{dll}{DLL}{Delay Locked Loop}
\newacronym{dmr}{DMR}{Dual Modular Redundancy}
\newacronym{dram}{DRAM}{Dynamic Random Access Memory}
\newacronym{dsd}{DSD}{Digital Synchronous Detection}
\newacronym{dsp}{DSP}{Digital Signal Processor}
\newacronym{dt}{DigiTime}{}
\newacronym{dvfs}{DVFS}{Dynamic Voltage and Frequency Scaling}
\newacronym{eda}{EDA}{Electronic Design Automation}
\newacronym{fdc}{FDC}{Frequency-to-Digital Converter}
\newacronym{fifo}{FIFO}{First In First Out}
\newacronym{fpga}{FPGA}{Field Programmable Gate Array}
\newacronym{gds}{GDS}{Global Data Share}
\newacronym{gnulgpl}{GNU LGPL}{GNU Lesser General Public Licence} 
\newacronym{gpgpu}{GPGPU}{General Purpose Graphics Processing Unit}
\newacronym{gpr}{GPR}{General Purpose Register}
\newacronym{gpu}{GPU}{Graphics Processing Unit}
\newacronym{gro}{GRO}{Gated Ring Oscillator}
\newacronym{io}{IO}{Input-Output}
\newacronym{hamsoc}{HAMSoC}{Hierarchical Agent Monitoring System-on-Chip}
\newacronym{hdl}{HDL}{Hardware Description Language}
\newacronym{hmp}{HMP}{Heterogeneous Multi-Processor}
\newacronym{ic}{IC}{Integrated Circuit}
\newacronym{icap}{ICAP}{Internal Configuration Access Port}
\newacronym[longplural={Intellectual Properties}]{ip}{IP}{Intellectual Property}
\newacronym{isa}{ISA}{Instruction Set Architecture}
\newacronym{lds}{LDS}{Local Data Share}
\newacronym{lru}{LRU}{Least Recently Used}
\newacronym{lsb}{LSB}{Least-Significant Bit}
\newacronym{lsu}{LSU}{Load Store Unit}
\newacronym{lut}{LUT}{Look Up Table}
\newacronym{mash}{MASH}{Multi-Stage Noise-Shaping}
\newacronym{mems}{MEMS}{Micro-Electro-Mechanical Systems}
\newacronym{miaow}{MIAOW}{Many-core Integrated Accelerator Of deepwater/Wisconsin}
\newacronym{mosfet}{MOSFET}{Metal Oxide Semiconductor Field Effect Transistor}
\newacronym{mpsoc}{MPSoC}{Multi-Processor System-on-Chip}
\newacronym{mshr}{MSHR}{Miss Status Holding/Handling Register}
\newacronym{noc}{NoC}{Network-on-Chip}
\newacronym{opencl}{OpenCL}{Open Computing Language}
\newacronym{ocn}{OCN}{On-Chip Network}
\newacronym{pcb}{PCB}{Printed Circuit Board}
\newacronym{pcie}{PCIe}{Peripheral Component Interconnect Express}
\newacronym{pl}{PL}{Programmable Logic}
\newacronym{pli}{PLI}{Verilog Programming Language Interface}
\newacronym{pll}{PLL}{Phase-Locked Loop}
\newacronym{ps}{PS}{Processing System}
\newacronym{pv}{PV}{Process Variation}
\newacronym{qoe}{QoE}{Quality of Experience}
\newacronym{qos}{QoS}{Quality of Service}
\newacronym{ram}{RAM}{Random Access Memory} 
\newacronym{risc}{RISC}{Reduced Instruction Set Computer}
\newacronym{riscv}{RISC-V}{Reduced Instruction Set Computing - V}
\newacronym{rtl}{RTL}{Register-Transfer Level}
\newacronym{sdk}{SDK}{Software Development Kit}
\newacronym{seec}{SEEC}{SElf-awarE Computing}
\newacronym{sgpr}{SGPR}{Scalar General Purpose Register}
\newacronym{si}{SI}{Southern Island}
\newacronym{simd}{SIMD}{Single Instruction Multiple Data}
\newacronym{simf}{SIMF}{Single Instruction Multiple Floating point}
\newacronym{sm}{SM}{Streaming Multiprocessor}
\newacronym{snr}{SNR}{Signal to Noise Ratio}
\newacronym[plural=SoCs,firstplural=Systems on Chip (SoCs)]{soc}{SoC}{System-on-Chip}
\newacronym{spared}{SPARED}{Self-aware PArtial Reconfiguration architecture for Edge Devices}
\newacronym{spice}{SPICE}{Simulation Program With Integrated Circuit Emphasis}
\newacronym{tad}{TAD}{Time \gls{adc}}
\newacronym[plural=TDCs,longplural={Time-to-Digital Converters}]{tdc}{TDC}{Time-to-Digital Converter}
\newacronym{tq}{TQ}{Time-Quantizer}
\newacronym{uart}{UART}{Universal Asynchronous Receiver/Transmitter}
\newacronym{vcdu}{VCDU}{Voltage Controlled Delay Unit}
\newacronym{vco}{VCO}{Voltage Controlled Oscillator}
\newacronym{vga}{VGA}{Video Graphics Array}
\newacronym{vhdl}{VHDL}{Very High Speed Integrated Circuit Hardware Description Language}
\newacronym{vlsi}{VLSI}{Very Large Scale Integration}
\newacronym{vgpr}{VGPR}{Vector General Purpose Register}
\newacronym{xilffs}{XILFFS}{Generic Fat File System Library}

\newacronym{amd}{AMD}{Advanced Micro Devices}
\newacronym{beol}{BEOL}{Back End Of Line}
\newacronym{cad}{CAD}{Computer-Aided Design}
\newacronym{cas}{CAS}{Circuits and Systems}
\newacronym{dfa}{DFA}{Discriminant Function Analysis} 
\newacronym{eu}{EU}{European Union}
\newacronym{fdd}{FDD}{fault detection and diagnostic}
\newacronym{fefet}{FeFET}{Ferroelectric Field Effect Transistor}
\newacronym{feline}{FeLINe}{FeFET Logic IN mEmory}
\newacronym[plural=FoMs,longplural={Figures of Merit}]{fom}{FoM}{Figure of Merit}
\newacronym{ga}{GA}{Genetic Algorithm}
\newacronym{hipeac}{HiPEAC}{High Performance and Embedded Architecture and Compilation}
\newacronym{hp}{HP}{Hewlett Packard}
\newacronym{hqp}{HQP}{Highly Qualified People}
\newacronym{hvac}{HVAC}{Heating, Ventilation and Air Conditioning}
\newacronym{ibm}{IBM}{International Business Machines corporation}
\newacronym{ict}{ICT}{Institute for Computer Technology}
\newacronym{iot}{IoT}{Internet of Things}
\newacronym{nda}{NDA}{Non-Disclosure Agreement}
\newacronym{nvp}{NVP}{Non-Volatile Processor}
\newacronym{oecd}{OECD}{Organization for Economic Cooperation and Development}
\newacronym{rd}{R\&D}{Research and Development}
\newacronym{rmsdd}{RMSDD}{Root-Mean Square of Successive Differences}
\newacronym{sdnn}{SDNN}{Standard Deviation Normal-to-Normal-Intervals}
\newacronym{soa}{SoA}{State-of-the-Art}
\newacronym{tsmc}{TSMC}{Taiwan Semiconductor Manufacturing Company}
\newacronym{tvlsi}{TVLSI}{Transactions on Very Large Scale Integration}

%% file: acronyms_local.tex
\usepackage[acronym]{glossaries} 

\newacronym{tmsl}{TMSL}{Three Memristor Stateful Logic}
\newacronym{uop}{$\mu$OP}{Micro-Operation}
\newacronym{rram}{RRAM}{Resistive Random Access Memory}
\newacronym{pwl}{PWL}{Piece-Wise Linear}
\newacronym{mvm}{MVM}{Matrix-Vector Multiplication}
\newacronym{felix}{FELIX}{Fast and Energy-efficient Logic}
\newacronym{fsm}{FSM}{Finite State Machine}
\newacronym{dag}{DAG}{Directed Acyclic Graph}
\newacronym{sl}{SL}{Stateful Logic}
\newacronym{ecm}{ECM}{Electrochemical Metallization}